\documentclass[10pt,a4paper,twocolumn]{article}

\usepackage[a4paper,
            left=0.75in, right=0.75in, top=0.75in, bottom=1in,
            columnsep=0.25in, nohead, nofoot]{geometry}

\usepackage[T1]{fontenc}
\usepackage{mathptmx}
\usepackage{textcomp}
\usepackage[final]{microtype}

\usepackage{graphicx}
\usepackage{amsmath}
\usepackage{array}
\usepackage{indentfirst}   % indent first paragraph after headings (template)
\makeatletter
\renewcommand\normalsize{%
  \@setfontsize\normalsize{10}{11.5}%
  \abovedisplayskip 5pt plus 2pt minus 2pt
  \belowdisplayskip \abovedisplayskip
  \abovedisplayshortskip 3pt plus 1pt
  \belowdisplayshortskip 3pt plus 1pt
  \let\@listi\@listI}
\normalsize
\makeatother
\makeatletter
\renewcommand\section{\@startsection{section}{1}{\z@}%
  {12pt plus 2pt minus 1pt}{1sp}%
  {\normalfont\fontsize{12}{13.8}\selectfont\bfseries}}
\renewcommand\subsection{\@startsection{subsection}{2}{\z@}%
  {11.5pt plus 2pt minus 1pt}{1sp}%
  {\normalfont\normalsize\bfseries}}
\makeatother

\usepackage[font=it, labelfont=it, labelsep=colon,
            justification=justified, singlelinecheck=false,
            skip=5pt]{caption}

\makeatletter
\renewenvironment{thebibliography}[1]{%
  \section{REFERENCES}%
  \list{\@biblabel{\@arabic\c@enumiv}}%
       {\settowidth\labelwidth{\@biblabel{#1}}%
        \setlength{\leftmargin}{0.3in}% [CHANGED] was 0.25in: room for two-digit labels ([10], [11])
        \setlength{\labelsep}{\dimexpr\leftmargin-\labelwidth\relax}%
        \setlength{\itemsep}{0pt}\setlength{\parsep}{0pt}%
        \setlength{\topsep}{0pt}\setlength{\partopsep}{0pt}%
        \usecounter{enumiv}%
        \let\p@enumiv\@empty
        \renewcommand\theenumiv{\@arabic\c@enumiv}}%
  \def\makelabel##1{##1\hfil}%
  \sloppy\clubpenalty4000\widowpenalty4000\sfcode`\.\@m}
  {\def\@noitemerr{\@latex@warning{Empty `thebibliography' environment}}%
   \endlist}
\makeatother

\newcommand{\um}{\,\textmu m}
\newcommand{\uW}{\,\textmu W}
\newcommand{\eij}[1]{\ensuremath{e_{#1}}}
\newcommand{\tabhead}[1]{\begin{tabular}[c]{@{}c@{}}#1\end{tabular}}% [CHANGED] two-line table headers (Table 1)
\newlength{\smallsigH}% [CHANGED] common image height for Figs. 4 and 5

\begin{document}

% ---------- Title / authors / affiliation (centered, both columns) ----
% No blank lines between title, authors and affiliation (template rule).
% All authors share one institution -> no superscripts (template rule).
\twocolumn[{%
  \centering
  {\fontsize{14}{16}\selectfont\bfseries
   \hyphenpenalty=10000\exhyphenpenalty=10000
   BILAYER PERIODICALLY POLED PIEZOELECTRIC \mbox{THIN-FILM} LITHIUM\\
   TANTALATE PLATFORM FOR \mbox{PASSIVE-LAYER-FREE} FLEXURAL AND\\
   TORSIONAL ACTUATION\par}
  {\fontsize{12}{13.8}\selectfont\itshape
   \mbox{Tzu-Hsuan Hsu}, \mbox{Jiajun Huang}, \mbox{Zihuan Liu},
   \mbox{Yen-Chen Wang},
   \mbox{Vakhtang Chulukhadze}, \mbox{Neal Hall}, \\ and \mbox{Ruochen Lu}\par}
  {\fontsize{12}{13.8}\selectfont
   The University of Texas at Austin, USA\par}
  \vspace{14pt}%
}]

% =====================================================================
\section{ABSTRACT}
This work demonstrates, to the best of our knowledge, the first passive-layer-free thin-film lithium tantalate (LT) actuator platform, supporting both out-of-plane flexural and in-plane torsional actuation from a single bilayer periodically poled piezoelectric film (P3F) X-cut LT stack. The polarity reversal across the bond interface flips the piezoelectric stress below the neutral axis, generating a bending moment without any passive elastic layer. Moreover, the two layers share identical elastic and thermal-expansion constants, making the stack thermally matched by construction. Suspended cantilever devices were fabricated on a single 2\,\texttimes\,1.5\um{} X-cut LT stack and verified by laser Doppler vibrometry (LDV). The tapered flexural resonator exhibits a 5.5\texttimes{} tip velocity amplification and a measured peak tip velocity of 10.3\,m/s at $Q=634$, and a 0.49\,MHz torsional cantilever resonator with $Q=514$ is demonstrated on the same stack. These results position bilayer P3F LT as a thermally matched, passive-layer-free actuation platform for vacuum-operated and optically interrogated MEMS such as scanners, gyroscopes, and ultrasonic transducers.

\section{KEYWORDS}
Lithium tantalate, periodically poled piezoelectric film, passive-layer-free
actuator, flexural resonator, torsional resonator, thermal mismatch.

% =====================================================================
\section{INTRODUCTION}

Out-of-plane flexural and in-plane torsional modes are key building blocks
of modern transducers. Flexural devices turn in-plane
piezoelectric strain into out-of-plane displacement for ultrasonic transducers~\cite{lu2020pmut} or inertial sensors~\cite{yazdi1998}. Torsional devices can provide angular motion for optical scanning mirrors~\cite{lu2023mirror} or rate gyroscopes~\cite{hodjat2020}.
In thin-film piezoelectric devices, the dominant structure of choice is the unimorph, in which an active piezoelectric film is paired with a passive elastic layer that offsets the neutral axis so that the piezoelectric stress produces a net bending moment~\cite{luo2021}.

The passive layer, however, has two main drawbacks. First, it adds mass without contributing to transduction. Second, its coefficient of thermal expansion (CTE) generally differs from that of the piezoelectric film~\cite{bartasyte2012}, so any temperature change bends the beam. This is particularly problematic in vacuum operation, where the absence of convective cooling turns even sub-\textmu W of absorbed heat from sources such as an optical probe beam into static bending.

Recently, P3F stacks with films of opposite polarity bonded together have replaced the passive layer in a unimorph to form a bimorph structure, as in lithium niobate (LN) piezoelectric micromachined ultrasonic transducers (PMUTs)~\cite{chulukhadze2026}, \cite{zhao2025}. In this work, a passive-layer-free flexural architecture is proposed in P3F LT. Compared with LN, LT offers a smaller temperature coefficient of frequency (TCF) in RF acoustics~\cite{kimura2019} and better optical properties~\cite{yu2024}. Here, we present the transduction principle of the bimorph P3F X-cut LT (X-LT) stack, paired with its CTE-matched-by-construction design validation, and transduction verification using five released devices by LDV, including four flexural resonators of different geometries and one torsional resonator.

\begin{figure}[!t]
  \centering
  \includegraphics[width=\columnwidth]{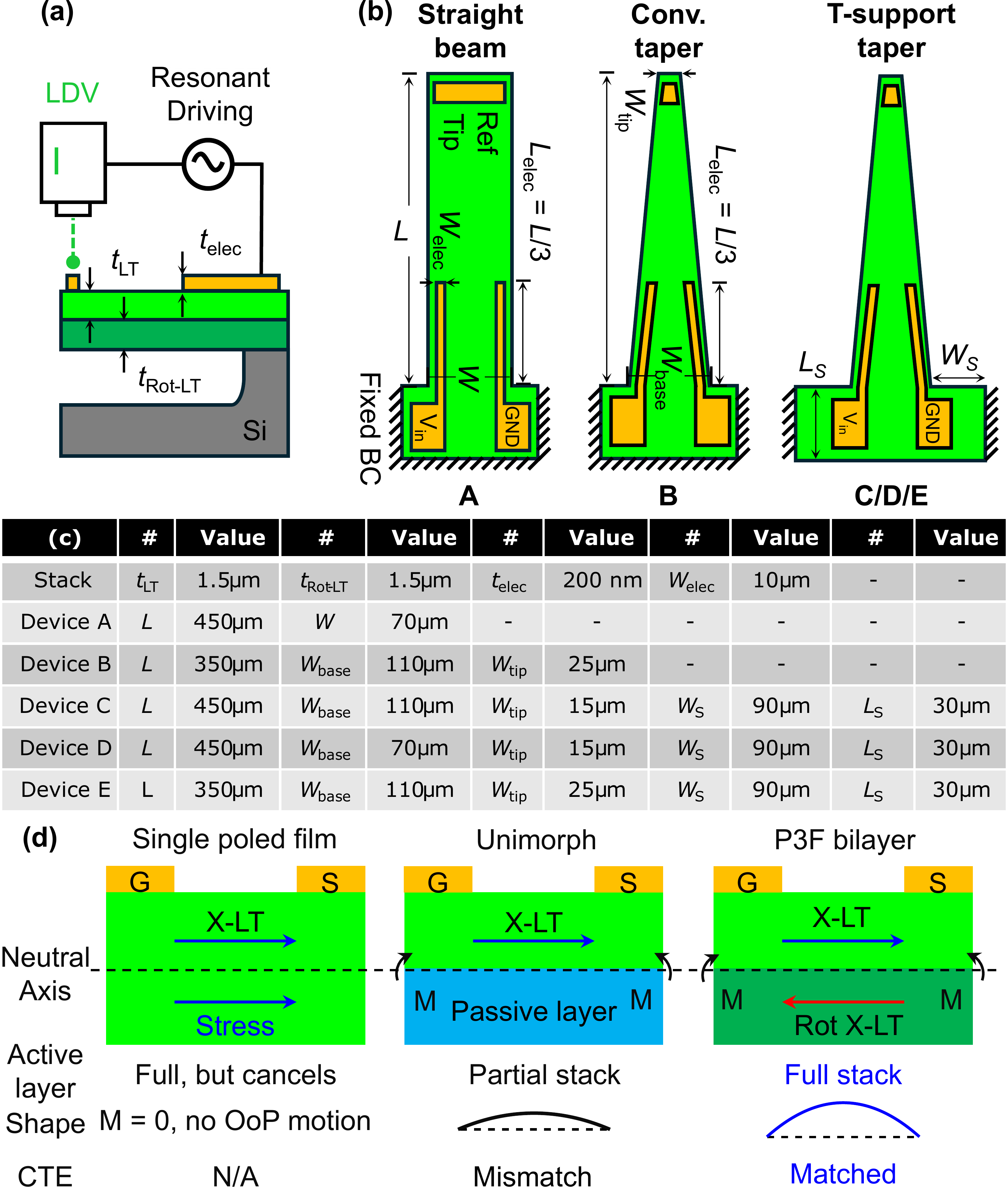}% [CHANGED] PDF figure
  \caption{(a) Side view of the bilayer P3F stack. (b) Top view of the
  three geometries covering five devices (A--E). (c) Stack and geometric
  information. (d) P3F principle: the polarity-inverted Rot-LT layer flips the stress below the neutral axis, producing a net bending moment.}
  \label{fig:stack}
\end{figure}

% =====================================================================
\section{DESIGN AND PRINCIPLE}
%\subsection{Bilayer P3F Stack and Device Geometry}
Fig.~1(a) shows the side view of the stack, which consists of two
1.5\um{} X-LT layers bonded with a 180\textdegree{} in-plane flip on high-resistivity Si, with 200\,nm Au
electrodes on top. The devices are driven through
electrodes ($V_\mathrm{in}$ and GND) of width
$W_\mathrm{elec}=10$\um{} running along the beam edges over
$L_\mathrm{elec}=L/3$ from the anchor with the device motion
measured by LDV. Fig.~1(b) shows three cantilever geometries covering five implemented 
designs, including a straight cantilever (Device~A), a tapered cantilever
(Device~B), and tapered cantilevers on T-shaped supports (Devices~C--E). The stack
and geometric parameters are listed in Fig.~1(c).

\begin{figure}[!t]
  \centering
  \includegraphics[width=\columnwidth]{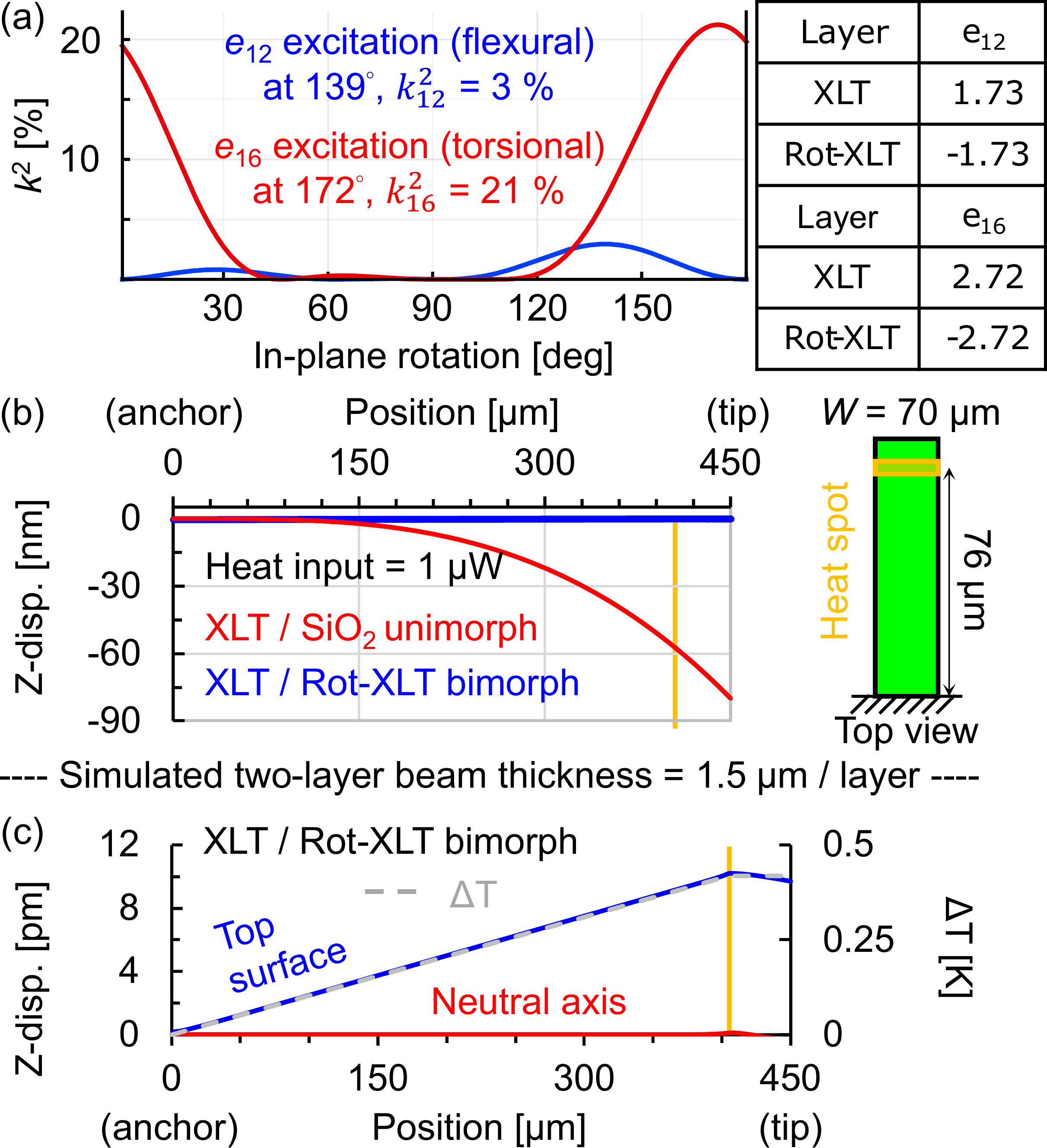}% [CHANGED] new Fig. 2 with panels (a)-(c)
  \caption{(a) Lateral-field coupling vs in-plane rotation of X-LT.
  (b) Thermal simulation of bimorph P3F film shows CTE-matched behavior.
  (c) Displacement profiles of the top surface and neutral axis of the bimorph P3F 
  film, showing linear expansion following $\Delta T$.}
  \label{fig:fem}
\end{figure}

%\subsection{Transduction Principle}
In this work, two accessible piezoelectric constants, \eij{12} and \eij{16}, were
selected under lateral-field excitation in X-LT to generate in-plane normal and shear
stress, respectively. Both stress fields are uniform through the film thickness
as illustrated in Fig.~1(d). Fig.~2(a) shows the calculated lateral-field coupling, defined as
\begin{equation}
  k_{1j}^2=\frac{e_{1j}^2}{\varepsilon_{11}^{S}\,c_{jj}^{E}},\qquad j=2,\,6,
\end{equation}
with all constants rotated to the in-plane device orientation. They peak at
139\textdegree{} ($k_{12}^2=3\%$) and 172\textdegree{} ($k_{16}^2=21\%$).
However, a uniform stress carries no moment about the neutral axis, so a
uniformly poled film extends rather than bends even though its full
thickness is active. A unimorph restores the moment by shifting the neutral
axis into a passive layer, at the cost of a partially active stack and a CTE
mismatch. The bimorph instead inverts the polarity below the neutral axis.
Considering $z=0$ at the LT-LT bonding interface, which is the neutral axis of two
elastically identical layers of thickness $t$, and neglecting the
electrodes, the piezoelectric moment per unit width under a lateral field
$E_1$ is
\begin{equation}
  M_\mathrm{p}=\int_{-t}^{t}\mathrm{sgn}(z)\,e_{12}E_1\,z\,\mathrm{d}z
  = e_{12}E_1\,t^{2},
\end{equation}
whereas the same integral vanishes for a uniformly poled film. An idealized
unimorph of the same total thickness with an elastically identical passive
layer yields only half of this moment at $e_{12}E_1t^{2}/2$ because only half of the 
stack is active. The same expression can also be applied to \eij{16} to give an
antisymmetric in-plane shear stress that generates the twisting moment for torsional 
motion.

Importantly, a 180\textdegree{} in-plane flip of X-LT about
the X axis reverses the sign of the piezoelectric constants but leaves the elastic,
dielectric, and thermal expansion constants unchanged. For a uniform temperature rise 
$\Delta T$, the thermal moment of the bilayer thus vanishes,
\begin{equation}
  M_\mathrm{th}=\int_{-t}^{t} Y\alpha\,\Delta T\,z\,\mathrm{d}z = 0,
\end{equation}
where $Y$ and $\alpha$ are the in-plane Young's modulus and CTE, identical
in both layers. The stack is thus CTE-matched by construction while
retaining full transduction efficiency.

% =====================================================================
\section{PHOTOTHERMAL SIMULATION}
Fig.~2(b) presents the thermal FEM simulation of a 450\um{} long, 
70\um{}
wide cantilever based on Device~A footprint with 1.5\um{} per layer 
under 1\uW{} of
absorbed power applied near the tip of the beam ($0.9L$). The X-LT/SiO$_2$
unimorph deflects by 80\,nm, while the bimorph shows no obvious 
bending. Fig.~2(c) zooms into the residual displacement of the 
bimorph; the picometer-level displacement is not bending. The 
neutral axis remains 
stationary while the top
surface follows the temperature-rise profile, corresponding to
through-thickness thermal expansion. The linear temperature rise 
from the
anchor to the heat spot reflects conduction along the beam as the 
only
heat path in vacuum. Consequently, the absorbed power
required to exceed a typical 2.5\um{} LDV depth of field rises from
32\uW{} for the unimorph to 245\,mW for the bimorph, an increase of
$\sim$7{,}700\texttimes{}.

% =====================================================================
\section{FABRICATION}
The LT is patterned by ICP-RIE to define the etch windows, followed 
by 200\,nm Au electrode deposition, and the devices are released by 
isotropic XeF$_2$ etching of the Si substrate. Fig.~3 shows the 
laser confocal 3D map and height profile of a
released device. The cantilever is flat within the confocal resolution
over a 60\um{} deep release cavity, with no residual-stress curl,
consistent with the symmetric bilayer stack observed in~\cite{kim2026}.

\begin{figure}[!t]
  \centering
  \includegraphics[width=\columnwidth]{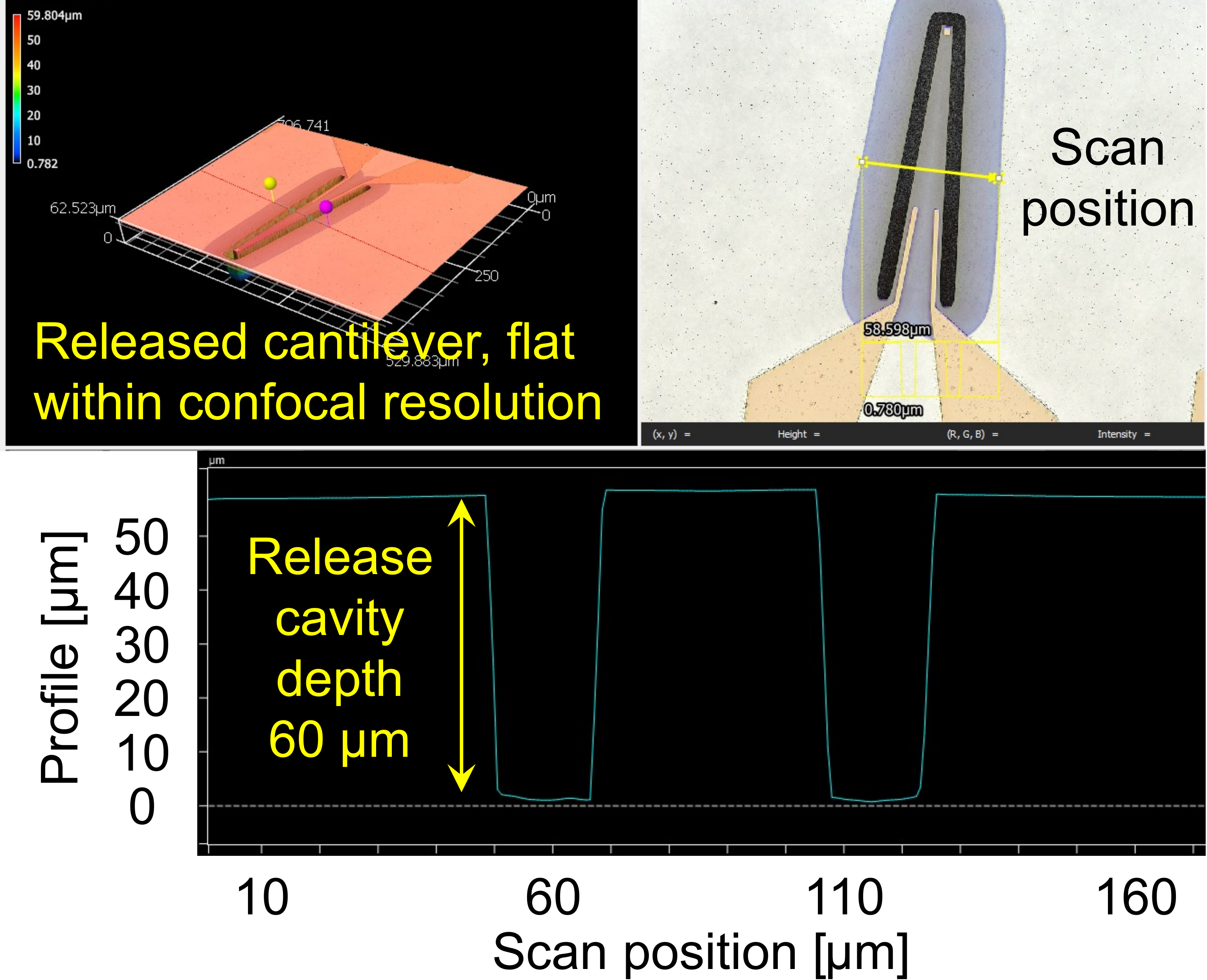}% [CHANGED] PDF figure
  \caption{Laser confocal 3D map and height profile of the released
  device.}
  \label{fig:confocal}
\end{figure}

\begin{figure*}[!t]
  \setlength{\smallsigH}{0.83\textwidth}% [CHANGED] was 0.665\textwidth (+25%) so the Conclusion continues onto p.4; set here, before the minipages
  \begin{minipage}[t]{\columnwidth}
    \centering
    \includegraphics[height=\smallsigH]{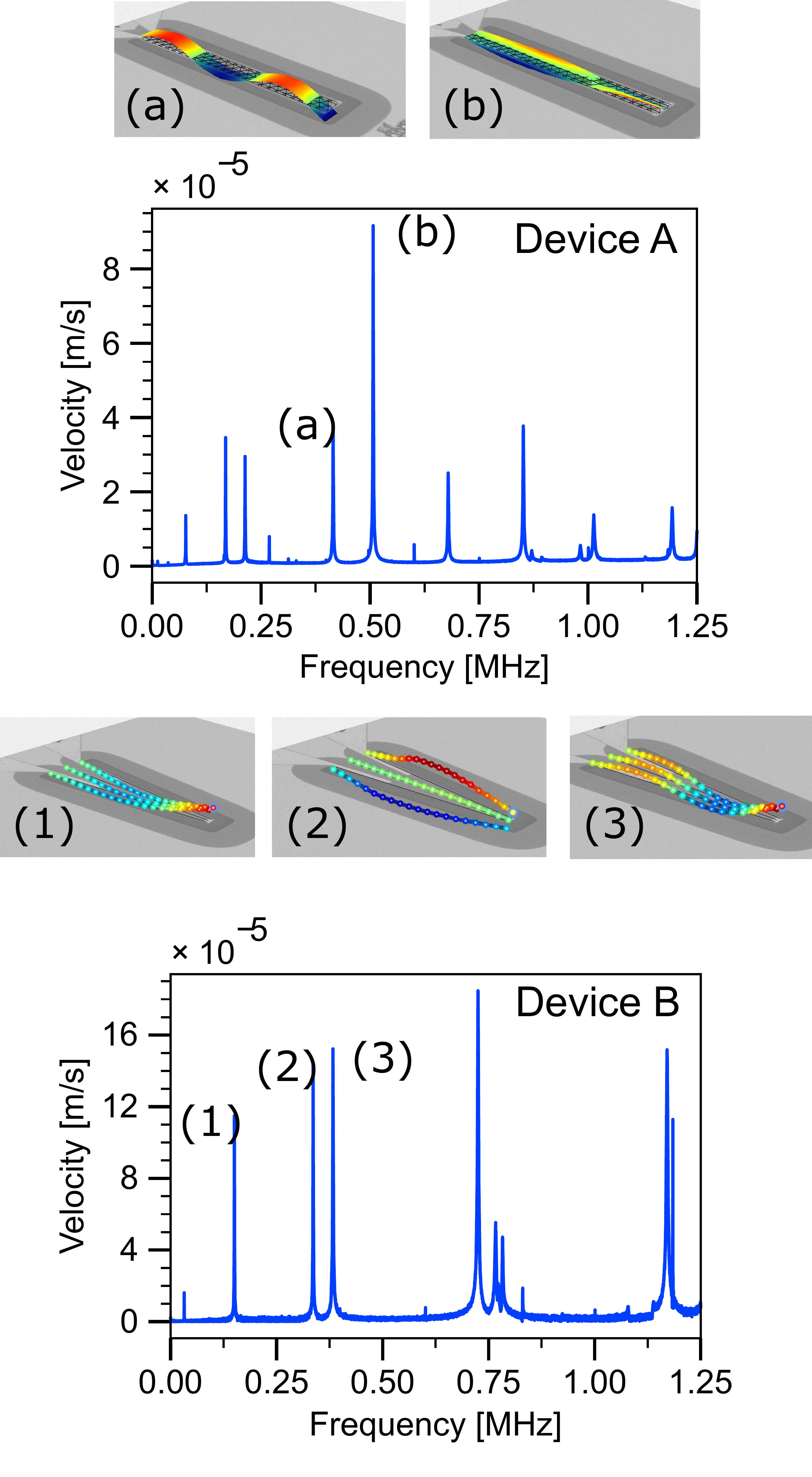}% [CHANGED] earlier size
    \caption{Small-signal LDV spectra and mode shapes of Devices~A and B.}
    \label{fig:smallAB}
  \end{minipage}\hfill
  \begin{minipage}[t]{\columnwidth}
    \centering
    \includegraphics[height=\smallsigH]{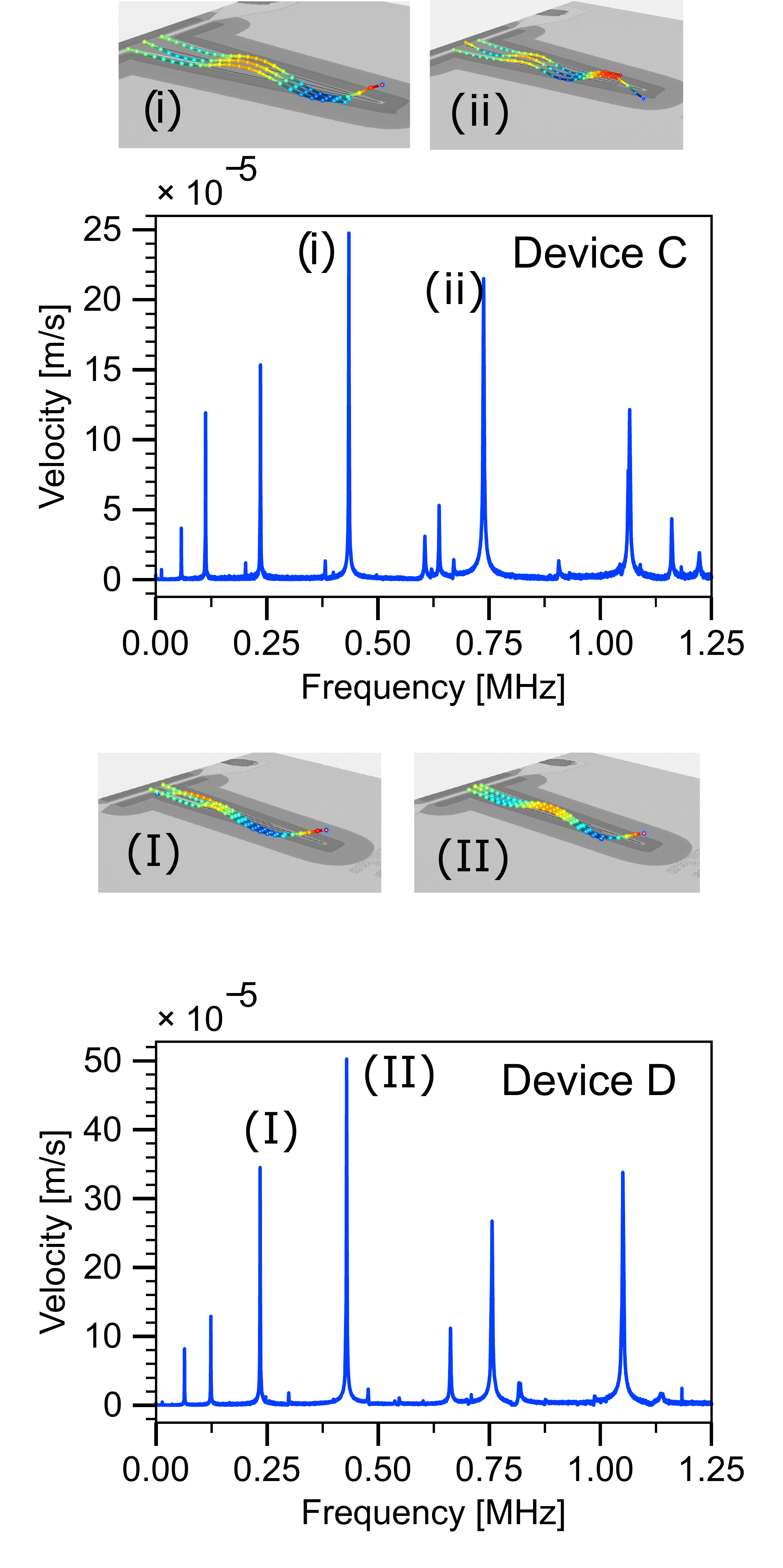}
    \caption{Small-signal LDV spectra and mode shapes of Devices~C and D.}
    \label{fig:smallCD}
  \end{minipage}
\end{figure*}

% =====================================================================
\section{EXPERIMENTAL RESULTS}
The small-signal LDV spectra of Devices~A/B and C/D were shown in Figs.~4 and~5, 
respectively. The devices were driven under broadband chirp excitation in 
air, and 
the measured mode shapes correspond to the labeled resonances. The extracted quality 
factors ($Q$) of the modes used for large-signal testing range from 443 to 634 and are summarized in Table~1. As all four devices share the same stack and electrode width, 
their difference in response is set by geometry. At identical drive, the tapered 
cantilever (Device~B) raises the peak velocity by $\sim$1.6\texttimes{} over the
straight beam design (Device~A) whereas the T-support tapered design (Device~C) raises  
the peak velocity by $\sim$2.7\texttimes{}. Further narrowing of the T-support base from 110 to 70\um{} (Device~D) boosts the peak velocity by $\sim$5.5\texttimes{}.

\begin{table}[!t]
  \caption{Summary of measured devices. Velocity factor is referenced to
  Device~A at identical small-signal drive.}
  \label{tab:summary}
  \centering
  \small
  \renewcommand{\arraystretch}{1.2}
  \setlength{\tabcolsep}{4pt}
  \begin{tabular}{|c|c|c|c|c|c|c|}
    \hline
    Device & $e_{ij}$ & \tabhead{Freq.\\ {[MHz]}} & $Q_\mathrm{3dB}$
    & \tabhead{Velocity\\ factor} & \tabhead{$V_\mathrm{max}$\\ {[m/s]}}
    & \tabhead{Limited\\ by} \\ \hline
    A & $e_{12}$ & 0.42 & 569 & 1\texttimes{} (ref)   & 2.1  & PA    \\ \hline
    B & $e_{12}$ & 0.38 & 484 & $\sim$1.6\texttimes{} & 4.7  & PA    \\ \hline
    C & $e_{12}$ & 0.73 & 443 & $\sim$2.7\texttimes{} & 6.8  & PA    \\ \hline
    D & $e_{12}$ & 0.43 & 634 & $\sim$5.5\texttimes{} & 10.3 & PA    \\ \hline
    E & $e_{16}$ & 0.49 & 514 & ---                   & 3.3  & Laser \\ \hline
  \end{tabular}
\end{table}

%\subsection{Large-Signal Response}
Figs.~6 and~7 show the corresponding large-signal time-domain velocity as the LDV drive signal is increased from 0.05\,V to 0.6--0.8\,V, amplified by a power amplifier (PA) with $\sim$640\texttimes{} gain. The maximum tested tip velocity rises from 2.1\,m/s for Device~A to 4.7, 6.8, and 10.3\,m/s for Devices~B, C, and D, respectively, with all devices surviving. The maximum velocity ceiling is currently limited by amplifier compression rather than by device failure.

\begin{figure}[t]
  \centering
  \includegraphics[width=0.31\textwidth]{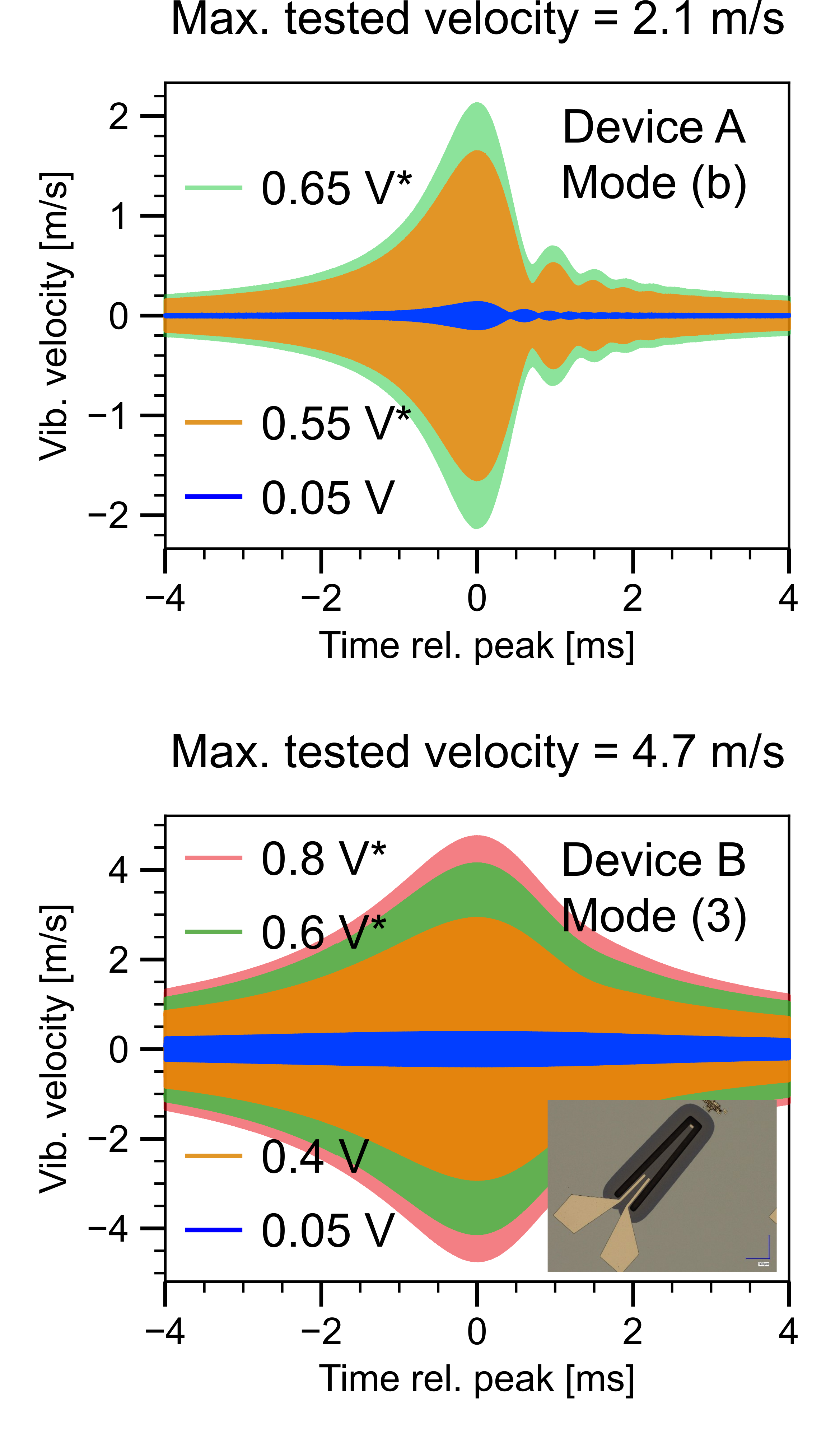}
  \caption{Time-domain velocity of Devices~A and B vs LDV drive signal. * Denotes 
  amplifier compression.}
  \label{fig:largeAB}
\end{figure}

\begin{figure}[t]
  \centering
  \includegraphics[width=0.31\textwidth]{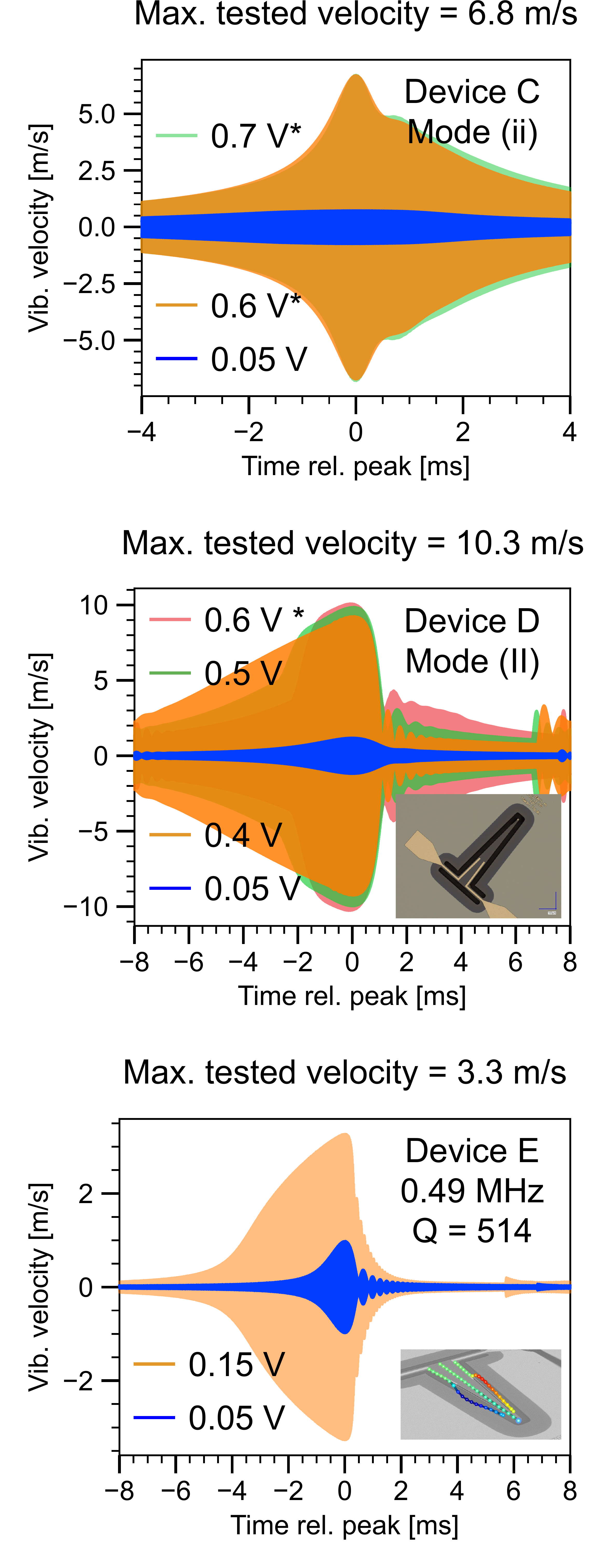}% [CHANGED] same width = same scale
  \caption{Time-domain velocity of Devices~C and D and of the torsional
  Device~E (\eij{16} excitation; inset: measured mode shape) vs LDV drive
  signal. * Denotes amplifier compression.}
  \label{fig:largeCDE}
\end{figure}

%\subsection{Torsional Actuation}
Aside from the flexural actuation demonstrated in this work, the torsional device 
(Device~E) using rotated in-plane \eij{16} excitation is also presented in Fig.~7. 
Here, the same stack supports torsional actuation at 0.49\,MHz
with $Q=514$. The velocity reaches 3.3\,m/s at a drive of only 0.15\,V. Unlike that of 
the flexural devices, the velocity of the torsional device is capped by optical dropout 
of the LDV signal under tilt at the edge of the cantilever.

% =====================================================================
\section{CONCLUSION}
A passive-layer-free actuator platform in bilayer P3F X-cut thin-film LT
has been demonstrated. Reversing the polarity across the bond interface
restores the bending moment that enables the bimorph actuation. We presented four 
flexural devices that reach a
5.5\texttimes{} geometric velocity amplification and up to 10.3\,m/s at
$Q=634$, and a torsional device that operates at 0.49\,MHz with $Q=514$ within the same 
stack. P3F LT is thus a promising multifunctional actuator platform for vacuum-operated 
sensing and optically interrogated MEMS.

% =====================================================================
\section{ACKNOWLEDGEMENTS}
The authors thank the DARPA NIMBUS program for funding support.

% =====================================================================

% =====================================================================
\section{CONTACT}
\begingroup\raggedright\parindent=0.25in
\indent*T.-H. Hsu; \mbox{tzuhsuan.hsu@austin.utexas.edu}\par
\endgroup

\end{document}